\documentclass[preprint,12pt]{elsarticle}

\usepackage{amssymb}
\usepackage{amsmath}
\usepackage{xurl}

\usepackage{lineno}

\journal{Applied Surface Science}

\begin{document}

\begin{frontmatter}



\title{Engineering Ferroelectric Phases in BaTiO$_3$ Films via Buffer-Layer-Induced Misfit Strain Reversal}


\author[1]{A.S. Mikheykin} %% Author name

\author[1]{N.V. Lyanguzov} %% Author name
\author[1]{A.S. Konstantinov\corref{cor1}}
\cortext[cor1]{Corresponding author}
\author[1,2]{V.B. Shirokov}
\author[1]{E.D. Linnik}
\author[1]{T. Bondar}
\author[3]{A.G. Razumnaya}
\affiliation[1]{organization={Southern Federal University},%Department and Organization
            city={Rostov-on-Don},
%            postcode={344090}, 
            country={Russia}}
            
\affiliation[2]{organization={Federal Research Centre The Southern Scientific Centre of the Russian Academy of Sciences},
    city={Rostov-on-Don},
    country={Russia}}

\affiliation[3]{organization={Jo\v{z}ef Stefan Institute},
    city={Ljubljana}, 
    country={Slovenia}}

%% Abstract
\begin{abstract}
Controlling ferroelectric phase symmetry through strain engineering is crucial for tailoring polarization topology and functional properties in oxide-based devices. Here, we show that a thin conductive buffer layer can override substrate-induced strain in BaTiO$_3$ (BTO) films on MgO.
Although the lattice parameter of MgO(001) would be expected to induce tensile misfit strain, in 500~nm BaTiO$_3$ films the thermal-expansion mismatch dominates, resulting in effective compressive stress upon cooling. Inserting a 50-nm La$_{0.67}$Sr$_{0.33}$MnO$_3$ (LSMO) buffer layer reverses the effective strain state from compressive to tensile. 
Using temperature-dependent polarized Raman spectroscopy, X-ray diffraction, and piezoresponse force microscopy, we reveal an unconventional phase sequence upon cooling: paraelectric → orthorhombic (\textit{aa}-phase, $\sim$407~K) → monoclinic (\textit{r}-phase, $\sim$300--345 K), characterized in-plane and tilted polarization components. The LSMO layer not only serves as a bottom electrode but also modifies the strain state, while interfacial charge redistribution induces a built-in electric field, evidenced by self-polarization and asymmetric PFM hysteresis. These findings highlight the critical role of conductive buffer layers in controlling ferroelectric symmetry and enabling oxide heterostructures with tailored polarization textures for low-power spintronics, tunable microwave devices, and neuromorphic computing.
\end{abstract}

\begin{graphicalabstract}
\includegraphics{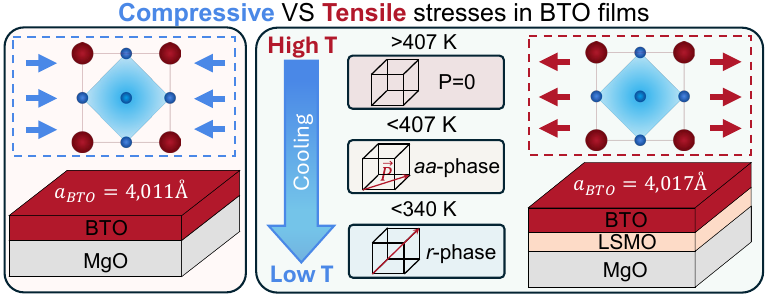}
\end{graphicalabstract}

\begin{highlights}
\item La$_{0.67}$Sr$_{0.33}$MnO$_3$ buffer modifies strain in BaTiO$_3$ films on MgO
\item Unconventional paraelectric $\rightarrow$  orthorhombic \textit{aa} $\rightarrow$  monoclinic \textit{r} transitions in BaTiO$_3$
\item Self-polarization and asymmetric piezoresponse loops indicate a built-in field
\end{highlights}

%% Keywords

\end{frontmatter}

%%\linenumbers

\section{Introduction}

Interest in ferroelectric materials for tunable microelectronic devices in the radio frequency range dates back to the mid-20th century \cite{Higa1956,Landauer1960,DiDomenico1962}. The development of ferroelectric thin-film technology has been driven by the ongoing demand for miniaturization and integration into existing microelectronic circuits. The strong anisotropy of ferroelectric materials makes epitaxial growth essential for thin films. A key consideration in epitaxial growth is the structural compatibility between the substrate and the deposited film, as well as the substrate's influence on the resulting film properties. This factor can be used to stabilize desired phases \cite{Janolin2009, Nguyen2011,Schlom2007,Haeni2004,Schlom2014,Guo2021} and microstructural features in thin films \cite{Guo2021,Schwarzkopf2017,Wang2019, Chapman2017} through the selection of the substrate material. Furthermore, integration with silicon substrates for microelectronic devices is achieved by depositing buffer layers of materials such as SrTiO$_3$ \cite{Spreitzer2021, Ngo2014}, yttria-stabilized zirconia \cite{Saanchez1998}, cerium oxide \cite{Nakagawara1995}, or  their combinations \cite{Saanchez1998,Nakagawara1995,Yamada2003,Kim2012}. In addition, for effective control of ferroelectric film properties, it is necessary to incorporate a conductive sublayer acting as a bottom electrode. Thin films with a perovskite structure, such as SrRuO$_3$ \cite{Kim2005}; La$_{1-x}$Sr$_x$MnO$_3$ \cite{Bea2006}; La$_{0.5}$Sr$_{0.5}$CoO$_3$ \cite{Wang2002} are often used for this purpose.

Among this class of conducting sublayers, one material of particular interest is: La$_{0.67}$Sr$_{0.33}$MnO$_3$. It exhibits colossal magnetoresistance (CMR) \cite{Ramirez1997}, meaning that its conductivity depends on an external magnetic field. Therefore, the heterostructure "ferroelectric - CMR film" can find application in spintronic devices, and a magnetic field provides an additional way to control device tunability.

Heterostructures composed of materials with different conductivities can induce effects that tune or modify the film properties. For example, the formation of a space-charge region at the interface creates a built-in electric field \cite{Karthik2013,Razumnaya2020}. The high electrostriction of polar perovskite oxides can lead to additional deformation, affecting the domain structure, direction, and magnitude of the dipole moment \cite{Li2014,Shirokov2015}. This can also play a vital role in the formation of a "dead" (non-switchable) ferroelectric layer in thin films \cite{Gagou2017}. Another possible effect of a built-in field is a significant shift in the critical temperatures of phase transitions, even in strain-free ferroelectric thin films \cite{Razumnaya2017}. For these reasons, studying interface effects in epitaxial heterostructures is of great importance for developing tunable devices based on ferroelectric thin films. However, the understanding of physical processes in multilayer heterostructures of ferroelectric films remain poorly understood.

In this work, we investigated the structure and dynamics of the crystal lattice of a 500-nm-thick BaTiO$_3$ (BTO) film deposited on magnesium oxide with a 50~nm-thick La$_{0.67}$Sr$_{0.33}$MnO$_3$ (LSMO) sublayer. We used Raman spectroscopy in the temperature range of 80–-500~K to determine the sequence and the temperatures of ferroelectric phase transformations in this heterostructure. The film is thick enough to minimize size effects and emphasize the role of interphase boundaries in influencing the properties of the BTO film. The experimental results are compared with the phenomenological theory of thin films of barium-strontium titanate, developed earlier to simulate the phase diagrams "concentration-misfit strain" \cite{Shirokov2009}. The effect of the built-in field on the structure and dynamics of the crystal lattice of a ferroelectric film is clearly shown.

\section{Methods}

\subsection{Synthesis}
A BTO film with a conductive LSMO buffer layer was grown on a cubic single crystal (001) MgO substrate by pulsed laser deposition (PLD). An excimer laser (KrF, 248~nm) with a laser energy density of 2.5 J/cm$^2$ and a pulse repetition rate of 10~Hz was used to deposit the heterostructure. A total of 650 and 10,000 laser pulses were used to deposit the buffer sublayer and BTO film, respectively. During deposition, the temperature of the substrate was maintained at 750~\textdegree C and the oxygen pressure inside the PLD chamber was 100~mTorr. The resulting heterostructure was annealed for 1 hour at 750~\textdegree C under an oxygen pressure of 300~mTorr.

\subsection{X-ray diffraction}
The crystallinity of the heterostructure, the preferred orientation, and unit cell parameters were determined using the X-ray diffractometer Rigaku Ultima IV. The XRD patterns of the sample were recorded using the Bragg-Brentano focusing geometry with filtered CuK$\alpha$ radiation.

\subsection{Piezoelectric force microscopy}

The surface of the BTO film was examined using an INTEGRA Therma atomic force microscope (NT-MDT). Piezoelectric properties were measured using a silicon cantilever with platinum coating and a buffer LSMO layer as the bottom electrode.

\subsection{Raman spectroscopy}
Temperature-dependent polarized micro-Raman spectra were obtained using Renishaw inVia Reflex spectrometer equipped with a NExT (Near-Excitation Tunable) filter to access low-frequency ranges down to 10~cm$^{-1}$. The micro-Raman spectra were excited using the polarized radiation from a solid-state laser ($\lambda$ = 514~nm) and collected in normal and side-view backscattering geometries with an optical analyzer. An optical microscope with a 50× objective was used to focus the incident light as a 2~$\mu$m-diameter spot on the sample. The temperature dependence of the Raman spectra was studied using a Linkam THMS600 heating/freezing stage, providing temperature stability of $\pm0.1$\textdegree C. The measured Raman spectra were corrected for the Bose-Einstein temperature factor. The polarized Raman spectra of the BTO/LSMO/MgO(001) thin film were obtained using the following laboratory Cartesian coordinate system with Z || [001]MgO.

\section{Results}

\begin{figure*}
	\centering
		\includegraphics[width=\linewidth]{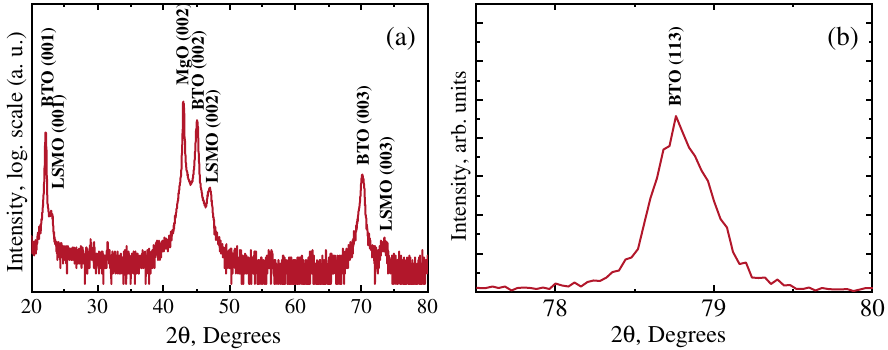}
	\caption{(a) Conventional XRD pattern for BTO/LSMO/MgO heterostructure. (b) Part of XRD pattern for tilted heterostructure. Miller indices are given in cubic representation of crystal structure.}
	\label{FIG:1}
\end{figure*}
The X-ray diffraction data at room temperature are shown in Figure \ref{FIG:1}. No impurity phases have been detected. Both the BTO film and the LSMO buffer layer are epitaxial. The out-of-plane unit cell parameter was determined from the conventional $\theta-\theta$ XRD pattern [Figure \ref{FIG:1}(a)] by applying a linear extrapolation using the Nelson-Riley function to the parameters calculated from each diffraction peak. The reciprocal space node with Miller indices (113) was profiled (see Figure \ref{FIG:1}(b)) to determine the in-plane unit cell parameter. The X-ray diffraction pattern treated under the assumption of tetragonal symmetry gives the following parameters of the pseudo-tetragonal unit cell: $c = 0.40304\pm0.00004$~nm, $a = 0.40178\pm0.00045$~nm, and $c/a = 1.003\pm0.004$.

\begin{figure*}
	\centering
		\includegraphics[width=\linewidth]{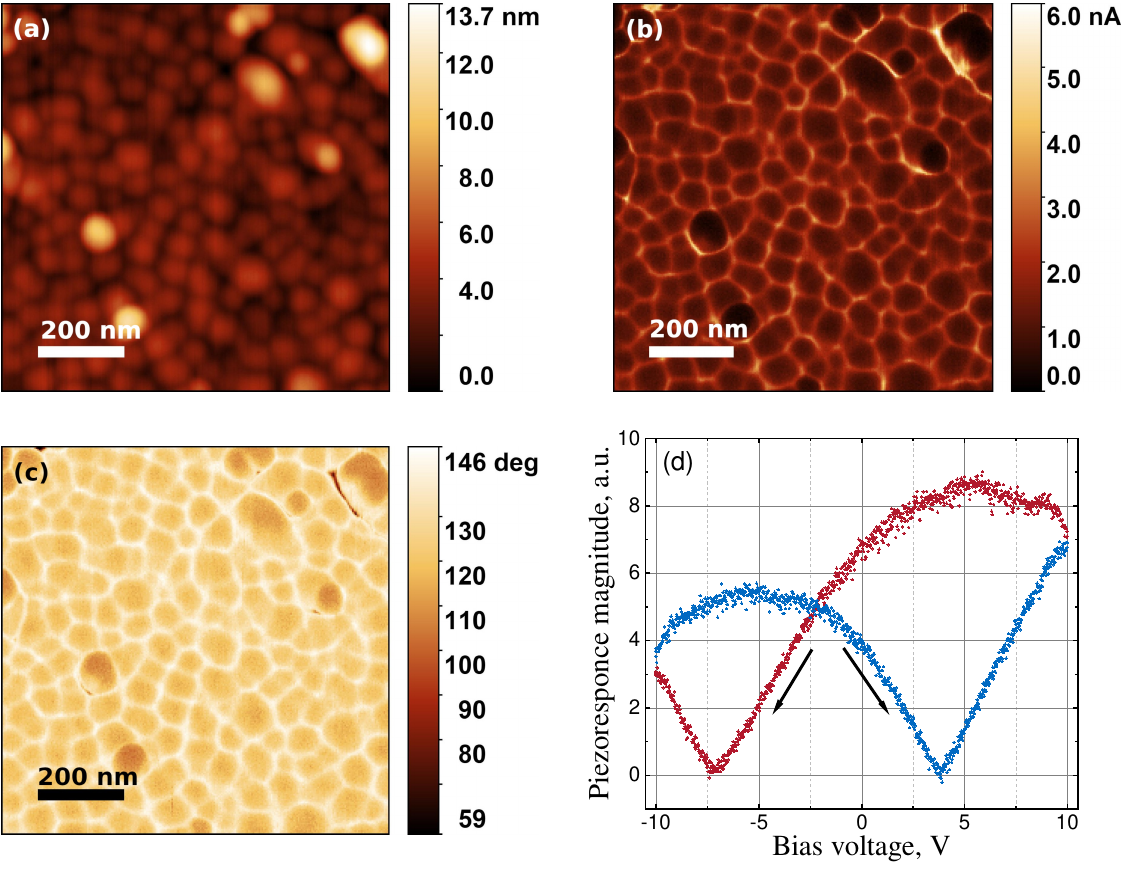}
	\caption{(a) Topography of surface; (b) magnitude and (c) phase of piezoelectric response; (d) PFM magnitude--voltage hysteresis loop measured at the center of a grain of the BaTiO$_3$/LSMO/MgO heterostructure.}
	\label{FIG:2}
\end{figure*}

Figure \ref{FIG:2}(a) shows the topography of a 1×1 $\mu~m^2$ surface fragment of the BTO/LSMO/MgO heterostructure. The surface morphology analysis showed that the film surface has a three-dimensional block structure with individual blocks smaller than 50~nm. The mean roughness and the mean square roughness of the heterostructure are 0.95~nm and 1.45~nm, respectively. 
Figures \ref{FIG:2}(b) and \ref{FIG:2}(c) show the magnitude of the piezoresponse and phase of the same region of the heterostructure surface, respectively. The domain structure of the sample is correlated with the size of the three‑dimensional blocks that constitute the barium titanate film and with the surface topography. Within each block, the polarization is uniform, as indicated by the phase map in Figure  \ref{FIG:2}(c). The as-grown BTO film exhibits unipolar polarization. The phase values on the piezoresponse distribution map indicate that the polarization in the barium titanate film is not oriented along the normal to the heterostructure surface. This suggests that the film's crystal structure is not tetragonal at room temperature, as is typical for thick films of barium titanate on a single-crystal MgO substrate without buffer layers \cite{ElMarssi2003}.
Figure \ref{FIG:2}(d) shows the PFM amplitude-voltage hysteresis loop on the BTO/LSMO/MgO heterostructure at the center of the grain.  Polarization switching processes occur at those voltages where the value of the piezoresponse amplitude drops to zero. The positions of the minima are at $-6.8$ and 3.8~V. The offset from zero may arise due to differences in the shape and material of the top and bottom electrodes when measuring the piezoresponse. The observed offset likely arises from a built-in constant electric field at the BTO/LSMO interface due to the difference in the electronic structure of these films \cite{Miao2016}. The presence of such a built-in field explains well the unipolar self-polarization of the thin film.

\begin{figure}
	\centering
		\includegraphics[scale=0.5]{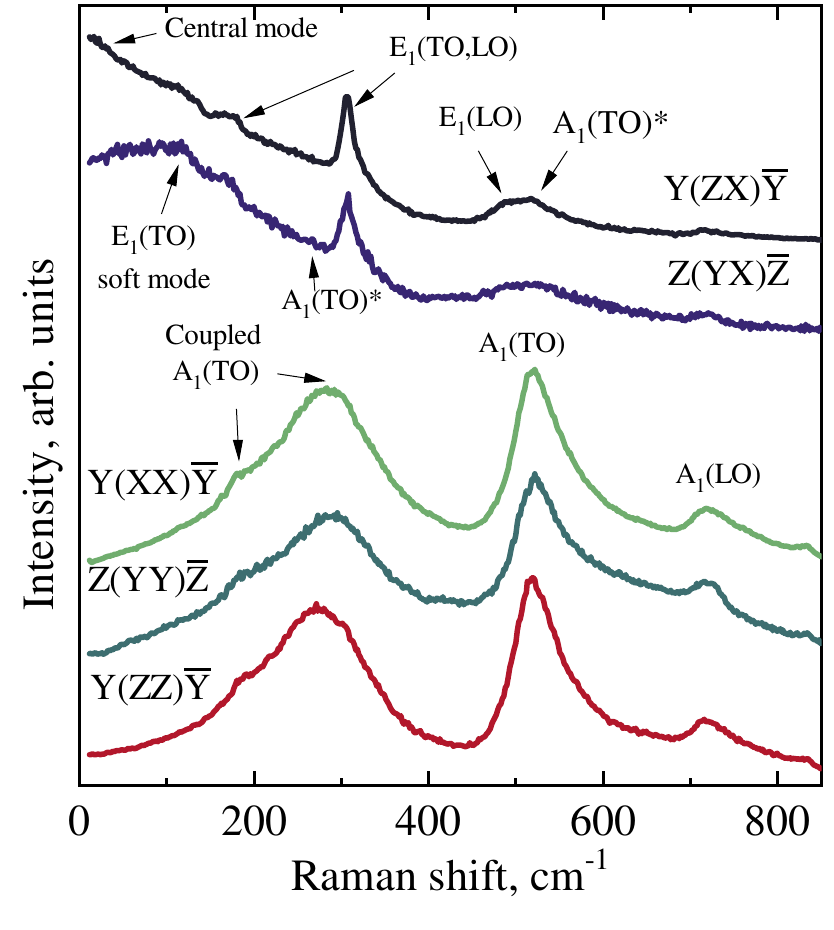}
	\caption{Polarized Raman spectra of the BTO film deposited on the MgO substrate with LSMO buffer layer obtained in diagonal (\textit{xx}, \textit{yy}, and \textit{zz}) and crossed (\textit{zx} and \textit{yx}) scattering geometries. All spectra were corrected for the temperature Bose-Einstein factor.}
	\label{FIG:3}
\end{figure}

Room-temperature polarized Raman spectra of the BTO film in diagonal and crossed scattering geometries are presented in Figure \ref{FIG:3}. 

The spectral response recorded in different scattering geometries contains all features typical of the tetragonal BTO crystal \cite{Scalabrin1977}, therefore, all modes were assigned according to the tetragonal symmetry with polarization oriented along the z-axis. In agreement with the Raman selection rules for a \textit{c}-domain film, the diagonal \textit{xx}, \textit{yy} and \textit{zz} spectra contain polar A$_1$ optical modes, whereas crossed \textit{xz} spectrum contains mainly polar modes with E-symmetry. In the tetragonal \textit{c}-phase, the low frequency E(TO) mode, so-called soft mode, corresponds to the polar displacements of Ti ions with respect to oxygen octahedra in a plane parallel to the substrate, while the A$_1$(TO) soft mode corresponds to displacements along the normal to the substrate \cite{Freire1988}. Additionally, in the ferroelectric phase, the A$_1$ and E modes exhibit splitting into transverse optical (TO) and longitudinal optical (LO) components due to the long-range interactions.

Although the spectra resemble those reported for a tetragonal BTO single crystal, we identified several important features indicating a more complex polarization configuration. In the diagonal \textit{zz} spectrum, no interference dip is observed, a feature typically present in the tetragonal phase at a frequency of $\sim$180~cm$^{-1}$. Furthermore, we observe the non-polar B$_1$ mode at $\sim$300~cm$^{-1}$ in the \textit{zz} spectrum, although this mode should be inactive in a \textit{c}-oriented film. According to the selection rules for the tetragonal \textit{c}-phase, no Raman-active modes are expected in the \textit{yx} geometry; however, an intense \textit{yx} spectrum is observed at room temperature. Moreover, Figure \ref{FIG:3} shows that the \textit{zx} and \textit{yx} spectra are similar; in addition to E-type modes, they contain pronounced, symmetry-forbidden A$_1$ bands centered at $\sim$250 and $\sim$540~cm$^{-1}$, which correspond to intense lines in the diagonal \textit{xx}, \textit{yy} and \textit{zz} scattering geometries. 

\begin{figure*}
	\centering
		\includegraphics[width=\linewidth]{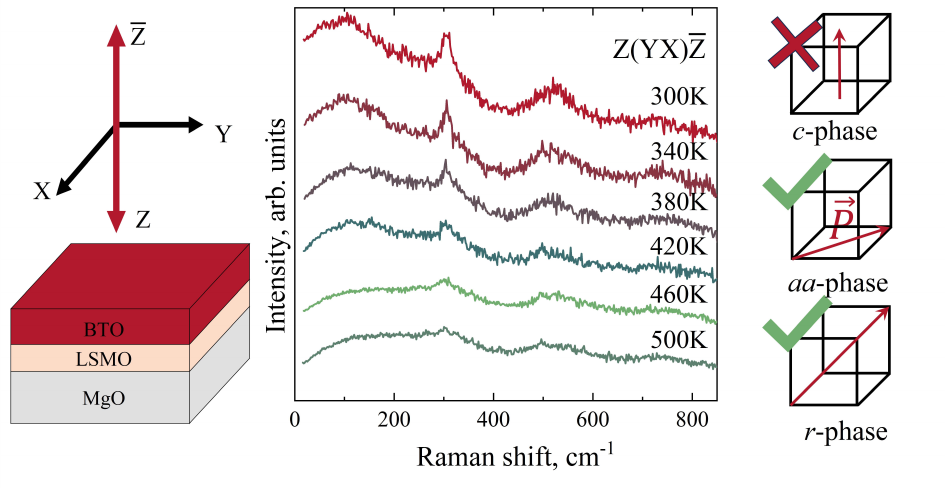}
	\caption{Raman spectra of the BTO/LSMO/MgO obtained in the crossed \textit{yx} scattering geometry. The left side shows Raman scattering geometry. The right side shows the possible polarization directions in the heterostructure.}
	\label{FIG:4}
\end{figure*}

Importantly, the Raman response in the \textit{yx} geometry persists up to 500~K, as shown in Figure \ref{FIG:4}, with well-defined phonon modes remaining clearly observable over the entire temperature range.

These observations suggest that our BTO film does not exhibit tetragonal symmetry with the polarization directed only along the \textit{z}-axis but instead possesses additional polarization components. 

At room temperature, according to phenomenological theory for films grown on the (001)-oriented cubic substrate \cite{Pertsev1998,Shirokov2007}, our BTO film may exhibit either monoclinic symmetry (\textit{r}-phase) with three non-zero polarization components ($p_x, p_y, p_z$) or orthorhombic symmetry (\textit{aa}-phase) with two in-plane components ($p_x, p_y, 0$), depending on the misfit strain. Accordingly, the possible structural states are effectively reduced to these non-\textit{c}-axis ferroelectric phases. However, due to the relaxation of Raman selection rules in strained thin films, the presence of domain structures, and possible local symmetry breaking at the BTO/LSMO interface, a definitive distinction between the \textit{aa} and \textit{r} phases cannot be made based on Raman data alone.

In our BTO film we observe a low-frequency central mode below 20~cm$^{-1}$ and a polar soft mode at $\sim$110~cm$^{-1}$ (Figure \ref{FIG:3}). Raman and far-infrared studies have established that the bulk BTO exhibits a complicated combination of order-disorder and displacive phase transitions \cite{Luspin1980}. Thus, in the ferroelectric phase, both the polar soft mode and Debye-like central modes can coexist in the Raman spectra, appearing in the sub-THz (1--30~cm$^{-1}$) and near-THz (30--100~cm$^{-1}$) frequency ranges. In addition to the soft mode, a similar Debye-type relaxation was also observed in BTO single crystals \cite{Petzelt1987, Hlinka2008}, ceramics, polycrystalline films \cite{Simon2010,Weerasinghe2013} and superlattices \cite{Tikhonov2015}. This central peak is associated with the large-amplitude relaxational motions of quasi-stable off-center Ti ionic positions.

\begin{figure*}
	\centering
		\includegraphics[scale=.75]{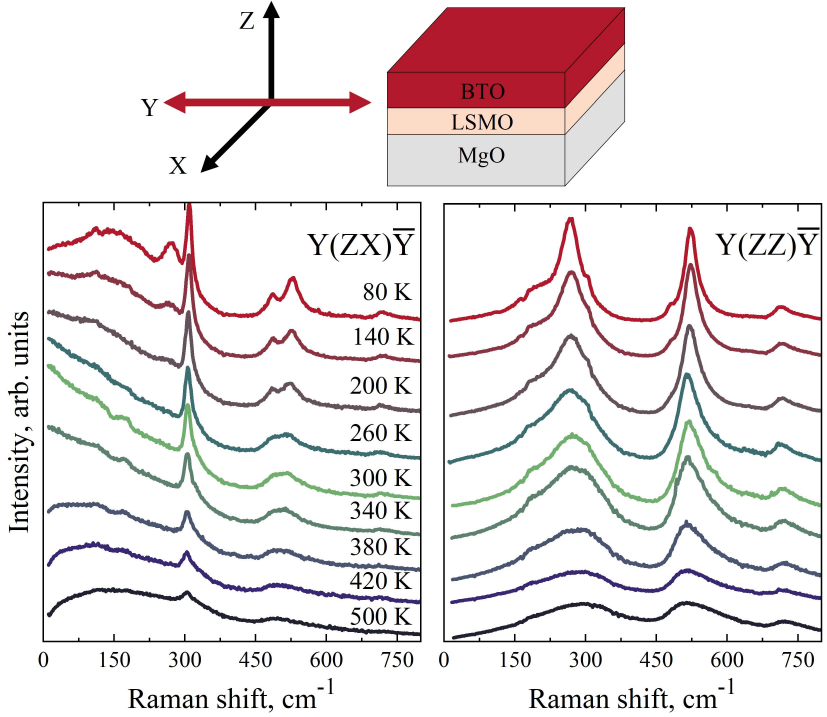}
	\caption{Raman spectra of the BTO/LSMO/MgO heterostructure obtained in the crossed zx and parallel zz scattering geometries at selected temperatures. All spectra were corrected for the temperature Bose-Einstein factor.}
	\label{FIG:5}
\end{figure*}

Figure \ref{FIG:5} presents the Raman spectra of the BTO/LSMO heterostructure in a temperature interval from 80 to 500~K for \textit{zx} and \textit{zz} scattering geometries. At high temperatures above 460~K, \textit{zz} spectra contain two weak broad bands at $\sim$230 and $\sim$580~cm$^{-1}$ attributed to the dynamic disorder of Ti ions in the paraelectric phase. Upon cooling, polar modes appear in both \textit{zx} and \textit{zz} spectra, with their intensity gradually increasing. The A$_1$ modes become narrower and more distinct in both geometries as the temperature decreases. The most notable changes occur in the low-frequency range (10--300~cm$^{-1}$), where the coexistence of the central and soft modes is observed in the \textit{zx} spectra between 140 and 420~K.

\begin{figure}
	\centering
		\includegraphics[scale=.9]{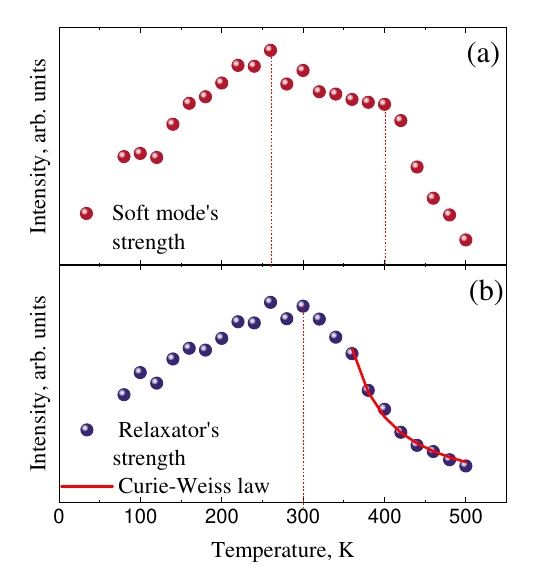}
	\caption{Temperature dependencies of (a) the soft mode strength and (b) the relaxator strength in BTO/LSMO heterostructure. The solid red line in panel (b) represents the theoretical fit using the Curie-Weiss law in the paraelectric phase}
	\label{FIG:6}
\end{figure}

Figure \ref{FIG:6} presents the temperature dependence of the strengths of the polar soft mode and the relaxator (central mode), extracted from the deconvolution of the \textit{zx} Raman spectra. As the temperature decreases from 500 K, both the soft mode strength [Fig. \ref{FIG:6}(a)] and the relaxator strength [Fig. \ref{FIG:6}(b)] progressively increase. In the high-temperature paraelectric state, the relaxator strength follows the Curie-Weiss law, which is well-fitted by the solid red line in Figure\ref{FIG:6}(b). A noticeable change in the slope for both curves occurs at $\sim$400 K, signaling the paraelectric-to-ferroelectric phase transition (from the cubic $0$-phase to the orthorhombic $aa$-phase). Upon further cooling, both parameters exhibit a broad maximum in the temperature range of 300–345 K, pointing to the subsequent structural transition into the monoclinic $r$-phase.

The combined Raman results indicate a sequence of symmetry changes in the BTO film upon cooling. The gradual polarization changes observed in the spectra and the temperature evolution of the soft and central modes point to a symmetry lowering from the paraelectric phase to the orthorhombic \textit{aa}-phase with polarization components ($p_x, p_y, 0$) at $\sim$400 K. Further cooling into the 300–-345~K range is accompanied by additional changes in the Raman response consistent with a transition to the monoclinic \textit{r}-phase with polarization components ($p_x, p_y, p_z$). 

For the precise interpretation of the Raman data, we use the thermodynamic theory developed for the BTO thin films grown on a single-crystal cubic substrate \cite{Shirokov2007}. For this, we must calculate the misfit strain of the studied BTO film and trace the thermodynamic path on the "misfit strain – temperature" ($u_m-T$) phase diagram. The misfit strain depends on temperature as follows \cite{Shirokov2007}:
\begin{equation} \label{eq:EQ1}
    u_m=u_0+(\alpha_0-\frac{b_0}{a_0}\alpha_s)(T-T_0)
\end{equation}
where $u_0$ is the misfit strain arising from the lattice mismatch between the film and substrate at the deposition temperature $T_0$, $a_0$ and $b_0$ are the lattice parameters of the film and substrate at room temperature respectively, $\alpha_0$ and $\alpha_S$ are the thermal expansion coefficients of the film and substrate respectively. According to \cite{Shirokov2007}, equation \ref{eq:EQ1} defines the thermodynamic path for the studied heterostructure on the $u_m-T$ phase diagram. The slope of the thermodynamic path on the phase diagram changes depending on the ratio of the lattice parameters of the film and substrate, and on the difference in their thermal expansion coefficients.

\begin{figure}
	\centering
		\includegraphics{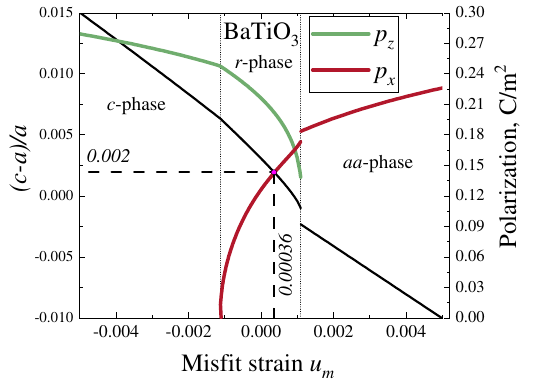}
	\caption{Components of the spontaneous polarization $p_x$ (solid red line) and $p_z$ (solid green line) and value of the tetragonality \textit{(c-a)/a} (solid black line) as functions of the misfit strain in the BTO thin film at T = 300 K. Dotted vertical lines indicate the phase boundaries at $u_m = -1.12\times10^{-3}$ and $u_m = 1.09\times10^{-3}$. Dashed lines indicate the $(c-a)/a$ value in the studied BTO film and to the corresponding $u_m = 0.36\times10^{-3}$.}
	\label{FIG:7}
\end{figure}

Figure \ref{FIG:7} shows the dependence of the tetragonality value \textit{(c-a)/a} and the components of the spontaneous polarization of the BTO thin film as a function of the misfit strain $u_m$ at room temperature. This plot was obtained numerically from equation \ref{eq:EQ2} presented in \cite{Shirokov2007}:
%% For 1 column view
\begin{equation} \label{eq:EQ2}
\begin{split}
        \frac{c-a}{a} = {} & \left(Q_{12}-\frac{s_{12}(Q_{11}+Q_{12})}{s_{11}+s_{12}}\right)(p^2_1+p^2_2) \\
        & + \left(Q_{11}-\frac{2s_{12}Q_{12}}{s_{11}+s_{12}}\right)p^2_3 - \frac{s_{11}-s_{12}}{s_{11}+s_{12}}u_m
\end{split}
\end{equation}
%% For 2 column view
%%\begin{equation} \label{eq:EQ2}
%%    \begin{split}
%%        \frac{c-a}{a}=\left(Q_{12}-\frac{s_{12}(Q_{11}+Q{12})}{s_{11}+s{12}}\right)(p^2_1+p^2_2)\\+\left(Q_{11}-\frac{2s_{12}Q{12}
%%        }{s_{11}+s_{12}}\right)p^2_3 - \frac{s_{11}-s{12}}{s_{11}+s_{12}}u_m
%%    \end{split}
%%\end{equation}
where \textit{c} and \textit{a} are the out-of-plane and in-plane lattice parameters  of the BTO film (as determined from XRD), $Q_{i,j}$, $s_{i,j}$ are electrostrictive coefficients, and the elastic compliances of the Landau potential in BTO film with cubic phase respectively, $p_i$ are the components of the spontaneous polarization.

According to x-ray measurements of our BTO film, \textit{(c-a)/a} is found to be 0.002, which corresponds to the misfit strain $u_m = 0.36\times10^{-3}$ in Figure \ref{FIG:7}. As can be seen, these parameters are attributed to the \textit{r}-phase with the polarization components $p_x$ and $p_z$ of 0.14 and 0.2~$C/m^2$ respectively.

\begin{figure}
	\centering
		\includegraphics{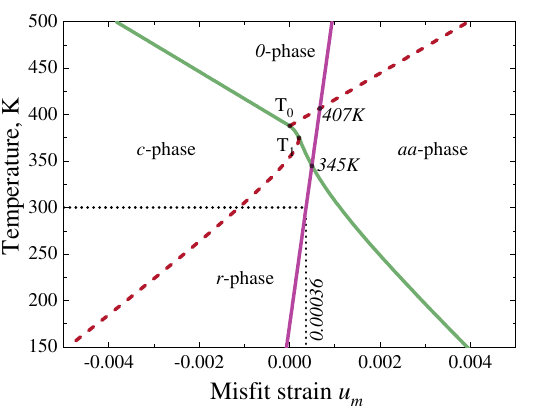}
	\caption{The "misfit strain-temperature" ($u_m -T$) phase diagram of the BTO thin film. The second-order phase transitions are denoted by red dashed lines. The solid green line indicates the first-order phase transition. The thermodynamic path of the BTO film grown on cubic (001)-oriented substrate with misfit strain $u_m = 0.00036$ is shown by the solid magenta line.}
	\label{FIG:8}
\end{figure}

The "misfit strain-temperature" ($u_m -T$) phase diagram for the BTO film grown on the cubic (001)-oriented substrate at room temperature calculated in \cite{Shirokov2007} from the phenomenological theory is shown in Figure \ref{FIG:8}. The thermodynamic path corresponding to the obtained $u_m = 0.36\times10^{-3}$ is presented in Figure \ref{FIG:8} taking into account the effect of the cubic MgO substrate. As can be seen from Figure \ref{FIG:8}, upon heating, the BTO film undergoes a first-order phase transition to the \textit{aa}-phase in the 300--345~K range, and then, at T = 407~K, the second-order phase transition to the paraelectric phase.

According to Raman data, above 400~K, we observe that the intensities of the polar modes decrease and the film undergoes a transition to the paraelectric phase. This transition temperature agrees with the calculated thermodynamic path.

\section{Discussion}
\subsection{Phase transitions sequence and misfit strain relaxation}
Polarized Raman spectroscopy of the 500~nm BTO film deposited on a single-crystal MgO substrate with a conductive 50~nm LSMO buffer layer revealed a sequence of phase transitions in the temperature range of 80--500~K.  Measurements in various scattering geometries, combined with the phenomenological theory, allowed us to identify the cooling sequence as a chain of \textit{0-}phase → \textit{aa-}phase → \textit{r-}phase transitions according to \cite{Pertsev1998} notation. This sequence differs from that of BTO films deposited directly on MgO without a sublayer, which follows \textit{0}-phase → \textit{c}-phase → \textit{r}-phase. 

The reason for this difference becomes clear if we consider the theoretical phase diagram "misfit strain -- temperature" for barium titanate films (see Fig. \ref{FIG:8} and Ref.\cite{Shirokov2007}). The cubic crystal lattice of MgO has a unit cell parameter of 4.21\AA, which means significant two-dimensional stretching during epitaxial growth of the BTO film. Perhaps this is the only widely used substrate that can create tensile stresses in barium titanate films.
However, for thick films, the stresses arising from lattice mismatch are partially relieved by the formation of mismatch dislocations, and the main factor influencing the final phase state becomes the mismatch in the thermal expansion coefficients of the film material and the substrate ($\alpha\text{(BTO)} = 10.4 \cdot 10^{-6} K^{-1}$ and $\alpha\text{(MgO)} = 14.8\cdot10^{-6} K^{-1}$). When the BTO/MgO structure cools, the substrate contracts significantly more, creating compressive thermoelastic stresses in the film. 
Therefore, all thermodynamic paths lie on the left side of the diagram for BTO/MgO heterostructures. 

\subsection{The role of the LSMO buffer layer}

It is precisely in the context of thick-film relaxation that the active role of interface engineering becomes evident. This study demonstrates that the introduction of a 50~nm-thick conducting LSMO sublayer between the BTO film and the MgO substrate reverses the sign of misfit strain relative to comparably thick BTO/MgO heterostructures \cite{ElMarssi2003}, effectively converting the two-dimensional compressive stress into tensile stress. 
The introduction of an LSMO buffer layer ($a \approx 0.387$~nm) onto an MgO substrate ($a \approx 0.421$~nm) results in tensile stress in the buffer layer. On the other hand, the buffer layer and the film are subjected to thermoelastic stresses from the substrate. Apparently, in the heterostructure under study, the competition between these mechanisms leads to the compensation of the compressive stresses observed in films without a sublayer.

The sequence of transitions for the BTO/LSMO/MgO heterostructure observed from the temperature dependence of the polarized Raman spectra indicates that the thermodynamic path for BTO film in such a composition of layers lies on the right side of the "misfit strain -- temperature" phase diagram. The introduction of a conductive LSMO buffer sublayer with a thickness of 50~nm is sufficient to shield the BTO film from compressive stress. It also reverses the sign of the misfit strain in the film as a result of the difference in lattice parameters between the buffer layer and the film. Thus, the LSMO film shows high flexibility, effectively decoupling the film from the substrate. 

\subsection{Raman spectroscopy and soft mode behavior under the built-in field}
According to Raman spectra, the transition from the nonpolar phase to the polar orthorhombic phase occurs at $\sim$407~K, and the transition from orthorhombic to polar monoclinic occurs in the 300--345~K range. The value of the misfit strain at room temperature, together with the values of the critical temperatures, allows us to conclude that the thermodynamic path has a slope in the "temperature--misfit strain" phase diagram (see Fig. \ref{FIG:8}). This slope can be attributed to two main reasons. First, there is a difference in the thermal expansion coefficients  of the film and substrate materials. Second, an internal electric field forms at the interface between the BTO film and the LSMO sublayer. The first mechanism is always present. Let us consider the presence of the contribution of the second mechanism using the example of the temperature behavior of polar modes in the Raman spectra of the BTO film.

\begin{figure*}
	\centering
		\includegraphics[width=\linewidth]{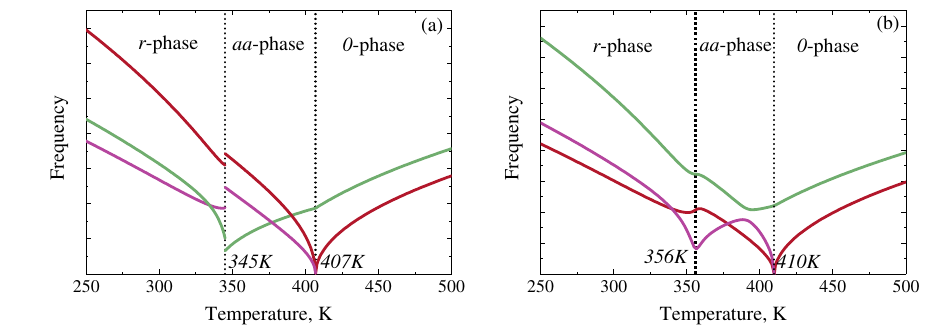}
	\caption{Temperature behavior of the soft mode frequencies calculated along the thermodynamic path of BTO films on a single-crystal MgO substrate with a conducting LSMO buffer layer (a) without an applied static electric field and (b) taking into account the built-in static field directed along the normal to the substrate surface.}
	\label{FIG:9}
\end{figure*}

The temperature behavior of soft modes can be a sign of the existence of a built-in field in the heterostructure, especially for the \textit{aa}-phase, where the in-plane component of the polarization vector of the film is significant. The parameters most sensitive to a static electric field are the frequency and intensity of the soft modes. This is confirmed by the known experimental results \cite{Fleury1967,Shigenari2015,Fleury1968,Steigmeier1968,Fleury1968a,Shigenari1981}, and by our calculations presented in Figure \ref{FIG:9}. 

Figure \ref{FIG:9}(a) shows the temperature dependence of the soft modes of the BTO film along the thermodynamic path realized for our heterostructure according to experimental data and theoretical concepts without taking into account the built-in electric field along the normal to the substrate surface. Figure \ref{FIG:9}(b) shows the temperature dependence of the frequencies of soft modes along the same thermodynamic path, but considering the existence of a static field along the normal to the substrate surface with a magnitude of $10^6$~V/m. The temperature-frequency dependence of the soft modes changes significantly under an applied electrostatic field. The most pronounced changes occur in the \textit{aa}-phase, in which the polarization lies in the plane of the substrate. It is clearly seen from the presented dependence that, in the temperature range of the existence of the \textit{aa}-phase, the soft‑mode frequencies exhibit a hill‑shaped temperature dependence. The temperature dependence of the soft mode intensity takes on a very similar form, which is confirmed by our experimental data. In addition, the Raman line at 300~cm$^{-1}$, which signals the presence of polar regions, remains intense across the entire temperature range studied. This feature can be attributed to the persistence of polar regions distributed throughout the film volume under the influence of the built-in field. Thus, the temperature dependences of the soft mode parameters obtained in this study, together with unipolar polarization in the  as-grown BTO film, allow us to conclude that a built-in field exists in the epitaxial BTO/LSMO/MgO heterostructure.

\begin{figure*}
	\centering
		\includegraphics[scale=1]{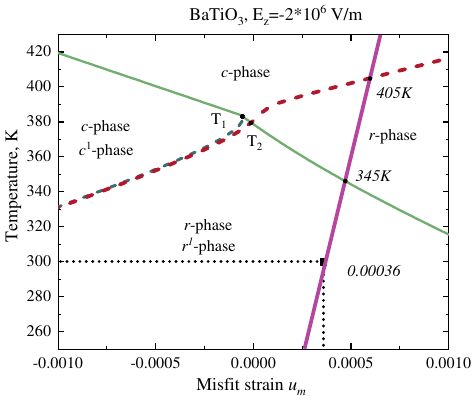}
	\caption{The "misfit strain-temperature" ($u_m -T$) phase diagram of the BaTiO$_3$ thin film under an out-of-plane electric field $E_z = -2 \cdot 10^5 \text{ V/m}$. Solid and dashed lines indicate first-order and second-order phase boundaries, respectively. The thermodynamic path of the BTO film grown on cubic (001)-oriented substrate with misfit strain $u_m = 0.00036$ is shown by the solid magenta line.}
	\label{FIG:10}
\end{figure*}

When comparing experimental data with the calculated phase diagram, it is important to take into account the effect of the built-in electric field. The presence of an out-of-plane field breaks the symmetry of the purely planar aa phase ($P_x, P_y, 0$), thereby inducing a vertical polarization component $P_z$. As can be seen from the calculated phase diagram for a film under an electric field of $E = -2 \cdot 10^6$ (see Fig. \ref{FIG:10}), Landau thermodynamic theory predicts that, the pure \textit{aa} phase vanishes, yielding to an \textit{r} phase within the temperature range of 345--405~K. Experimentally, however, data regarding the relaxor and the soft mode strengths reveal twwo successive transitions are observed at $\sim$407~ K and $\sim$300--340~ K. Consequently, a pseudo-\textit{aa} phase persists in the 340--407~K range, while the true \textit{r} phase emerges only below $\sim$340~ K.

This demonstrated capacity of the buffer layer to alter the strain state highlights a subtle point that is often oversimplified in the broader literature. For instance, an extensive review of the literature, presented as a table of “film material -- substrate” combinations, was included as an appendix in the 2007 book Physics of ferroelectrics: a modern perspective \cite{Rabe2007}. 
In this compilation, as in the original papers, the primary parameter controlling the misfit strain was assumed to be the lattice mismatch between the film and the substrate. This was the case even for studies on BaTiO$_3$, where a conducting SrRuO$_3$ buffer layer was often used. The Sr--Sr distance is 3.81\AA\ in bulk strontium ruthenate at room temperature \cite{Cuffini1993}, which sets the unit cell parameter in the cubic representation of the perovskite structure. This means significant compressive two-dimensional stresses in the barium titanate film, which, according to the review's authors, were determined only by the substrate material. 
It is evident from our findings that the conducting buffer layer in the analysis of combinations of "film material -- substrate" cannot be regarded only as a lower electrode that does not affect the phase states and properties of thin films. Of course, our result does not indicate that everything was previously misinterpreted. However, considering the results obtained, it is necessary to revise the already accumulated experimental data and to conduct a more detailed study of the effect of conducting or matching layers on the state and properties of ferroelectric thin films.

\section{Conclusion}

PFM and polarized Raman spectroscopy reveal that the BTO/LSMO/MgO heterostructure does not exhibit tetragonal symmetry at room temperature, contrary to expectations. Temperature-dependent Raman spectra show a paraelectric-ferroelectric transition near 407~K, from the \textit{0}-phase to the \textit{aa}-phase with in-plane polarization, which then transforms to the \textit{r}-phase with three non-zero polarization components in the 300--345~K range. This behavior contrasts with that of BTO films grown on MgO(001) substrates, which experience tensile misfit strain. Introducing a thin buffer layer effectively shields the BTO film from the tensile stresses of the substrate.  The significant impact of such a thin layer on a comparatively thick film raises questions about the mechanisms of strain transfer and polarization control in ferroelectric heterostructures, suggesting that systematic studies of the buffer layer effects could provide new approaches to phase and properties engineering. 

\section*{Acknowledgment}
This research was carried out with the support of the Russian Science Foundation (project No. 25-22-00478) at the Southern Federal University.

%% Use figure environment to create figures
%% Refer following link for more details.
%% https://en.wikibooks.org/wiki/LaTeX/Floats,_Figures_and_Captions
%:\begin{figure}[t]%% placement specifier
%% Use \includegraphics command to insert graphic files. Place graphics files in 
%% working directory.
%\centering%% For centre alignment of image.
%\includegraphics{example-image-a}
%% Use \caption command for figure caption and label.
%\caption{Figure Caption}\label{fig1}
%% https://en.wikibooks.org/wiki/LaTeX/Importing_Graphics#Importing_external_graphics
%\end{figure}

\
\bibliographystyle{elsarticle-num} 
\bibliography{BTO-LSMO}

\end{document}